\documentclass[10pt,a4paper,preprint,aps,nofootinbib,floatfix]{revtex4-1}

\usepackage{dcolumn}
\usepackage{bm}
\usepackage{graphicx}
\usepackage{caption}
\usepackage{subcaption}
\usepackage{amsmath}
\usepackage{natbib}
\usepackage{enumitem}

\begin{document}

\title{Predicting language diversity with complex networks}
\date{\today}

\author{Tomasz Raducha}
\email{tomasz.raducha@fuw.edu.pl}
\affiliation{Institute of Experimental Physics, Faculty of Physics, University of Warsaw, Pasteura 5, 02-093 Warsaw, Poland}
\author{Tomasz Gubiec}
\affiliation{Institute of Experimental Physics, Faculty of Physics, University of Warsaw, Pasteura 5, 02-093 Warsaw, Poland}
\affiliation{Center for Polymer Studies, Boston University, Boston, MA 02215 USA}


\maketitle



\textbf{
Evolution and propagation of the world's languages is a complex phenomenon,
driven, to a large extent, by social interactions
\cite{beckner2009language,mufwene2002competition,tomasello2010origins}.
Multilingual society can be seen as a system of interacting agents
\cite{eckert2000language,baxter2009modeling,carro2016coupled},
where the interaction leads to a modification of the language spoken by the individuals
\cite{lieberman2007quantifying,bybee2006usage}.
Two people can reach the state of full linguistic compatibility due to the positive interactions,
like transfer of loanwords.
But, on the other hand, if they speak entirely different languages, they will separate from each other.
These simple observations make the network science \cite{albert2002statistical} the most suitable framework
to describe and analyze dynamics of language change
\cite{castello2008modelling,schulze2008birth,hruschka2009building}.
Although many mechanisms have been explained
\cite{abrams2003linguistics,loreto2007social,patriarca2012modeling,sutherland2003parallel},
we lack a qualitative description
of the scaling behavior for different sizes of a population.
Here we address the issue of the language diversity in societies of different sizes,
and we show that local interactions are crucial to capture characteristics of the empirical data.
We propose a model of social interactions, extending the idea from \cite{raducha2017coevolving},
that explains the growth of the language diversity
with the size of a population of country or society.
We argue that high clustering and network disintegration are
the most important characteristics of models
properly describing empirical data.
Furthermore, we cancel the contradiction between previous models
\cite{axelrod1997dissemination,sanmiguel2007}
and the Solomon Islands case.
Our results demonstrate the importance of the topology of the network,
and the rewiring mechanism in the process of language change.
}



\begin{figure}
\centering
  \begin{subfigure}{0.15\textwidth}
  \centering
  \includegraphics[width=1.0\linewidth]{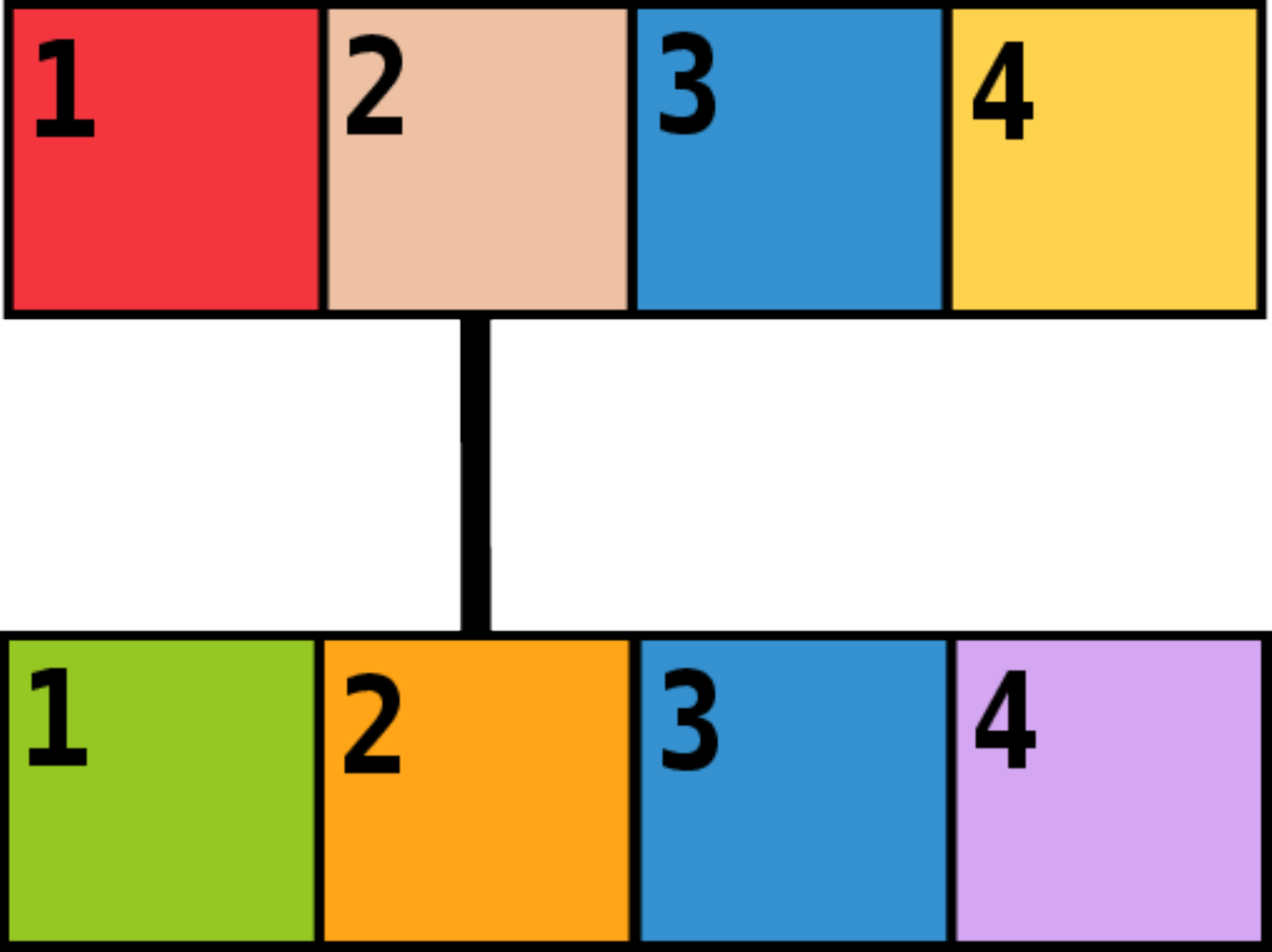}
  \caption{}
  \end{subfigure}
  \hspace{2cm}
  \begin{subfigure}{0.15\textwidth}
  \centering
  \includegraphics[width=1.0\linewidth]{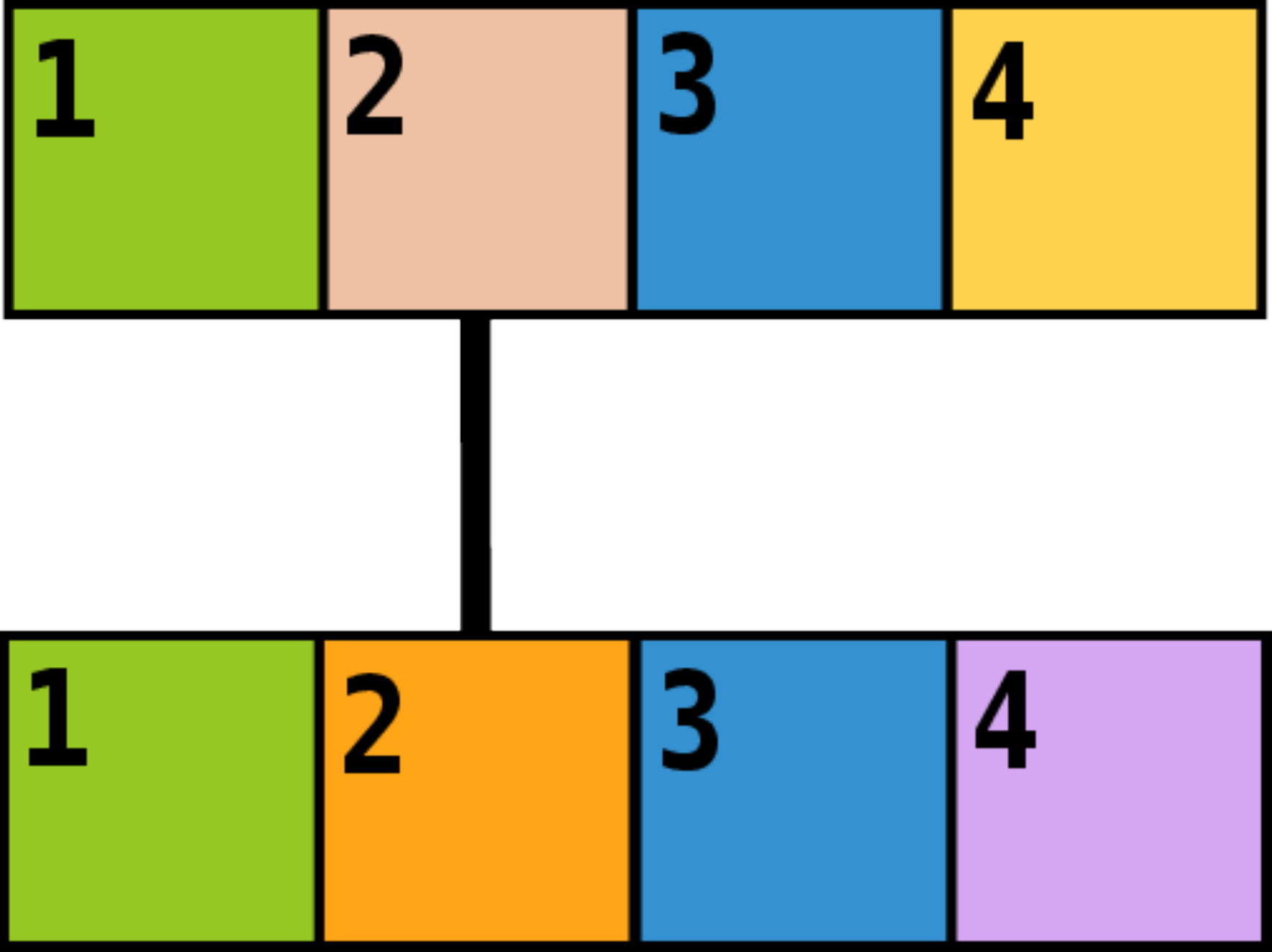}
  \caption{}
    \end{subfigure}
\caption{Schematic illustration of the positive interaction for $F=4$.
Bars illustrate four traits (from 1 to 4) of two interacting nodes. Every color is one of $q$ possible values.
The upper node is the active one. (a) Before interaction -- trait number 3 has the same value (blue).
(b) After interaction -- in one of the possible realizations the active node adopts value of the first trait (green).
\label{fig:interaction}}
\end{figure}

\begin{figure}
\centering
  \begin{subfigure}{0.3\textwidth}
  \centering
  \includegraphics[width=1.0\linewidth]{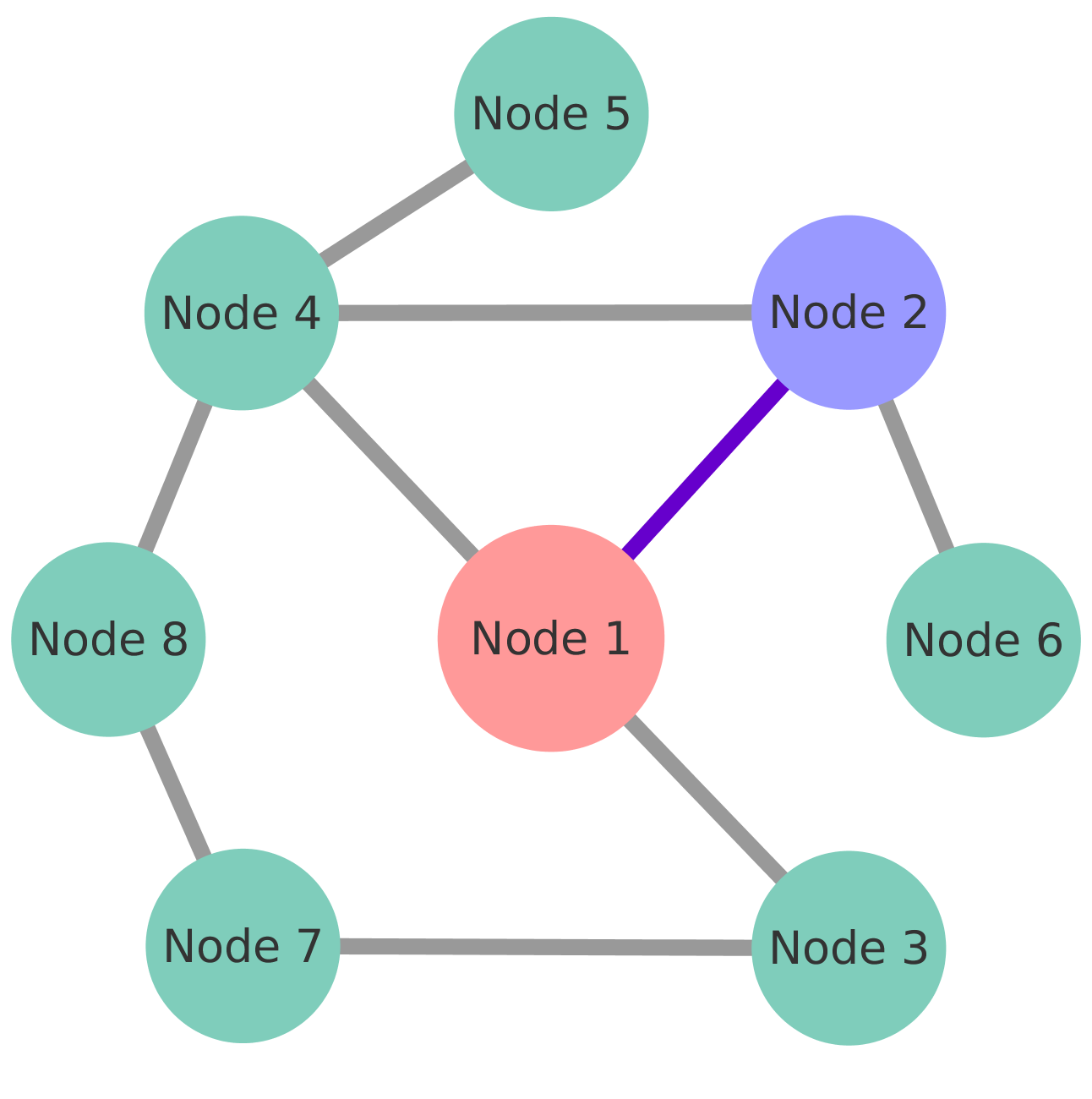}
  \caption{}
  \end{subfigure}
  \hspace{2cm}
  \begin{subfigure}{0.3\textwidth}
  \centering
  \includegraphics[width=1.0\linewidth]{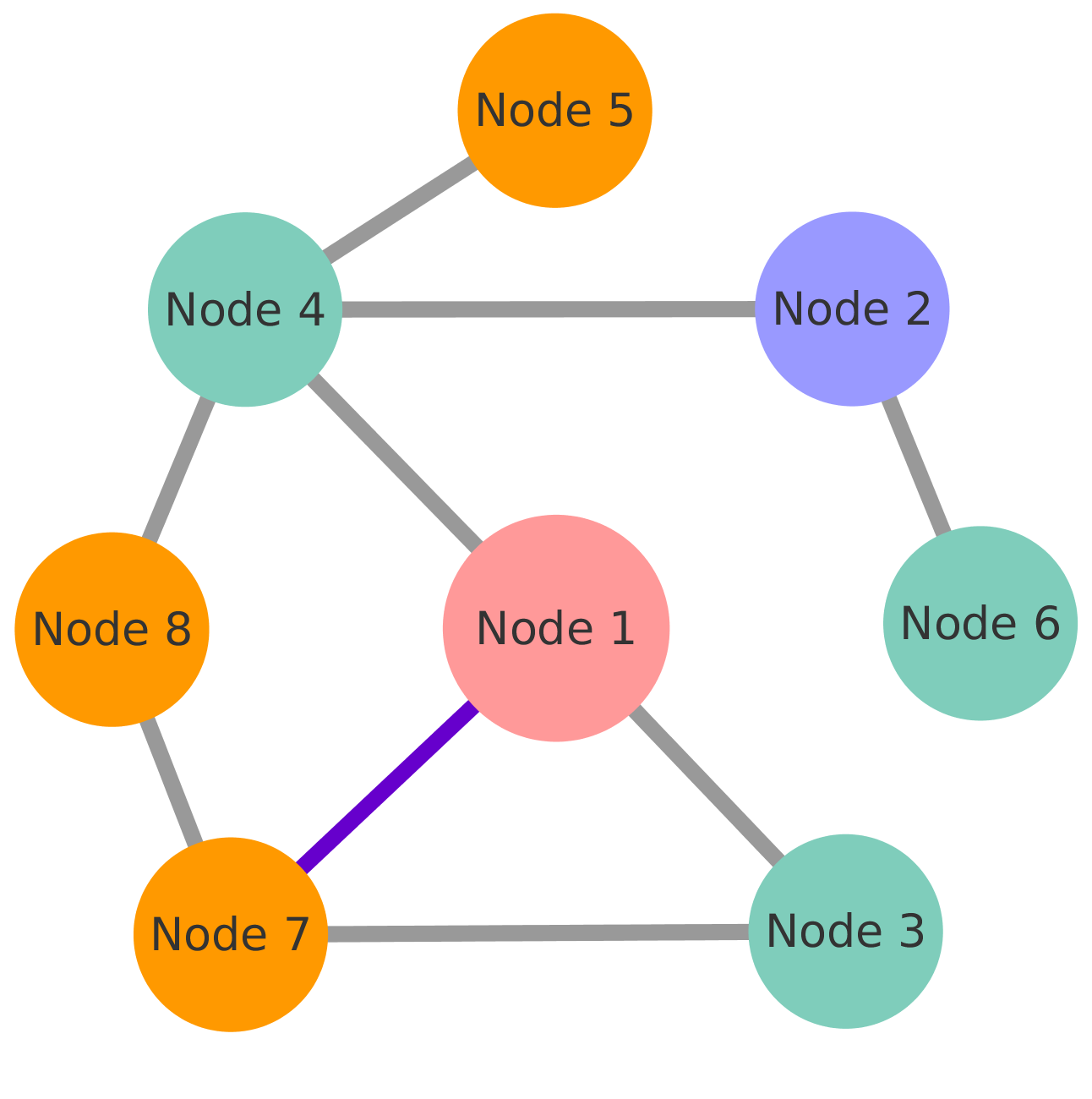}
  \caption{}
    \end{subfigure}
\caption{Schematic illustration of the rewiring mechanism.
Node 1 (pink) is the active node. (a) Consider its interaction with node 2 (purple).
Assuming they have no common traits, the link between them must be rewired.
(b) After erasing the edge, the active node can create a link to one of the nodes 5, 7, or 8 (orange ones).
Node 7 is randomly selected.
\label{fig:rewiring}}
\end{figure}

Consider a system of $N$ individuals, each using language described by a set of $F$ traits.
Individuals are connected by links, indicating social interactions enabling language transfer.
Two agents can speak very similar dialects or completely different languages, what is reflected
in $q$ different values of every trait.
Traits should be interpreted as groups of words, or grammar rules, rather than single words.
During the interaction people tend to adapt their languages
to each other, if they have anything in common (see FIG. \ref{fig:interaction}).
The more similar languages they speak, the more
probable is the positive interaction and learning from each other, leading to a further increase of the similarity.
On the other hand, people using languages with all traits different have no possibility
to communicate and will cut the connection and look for a new neighbor.
After disconnecting from a neighbor, active node will choose a new one from
a set of vertices distant by two edges (see FIG. \ref{fig:rewiring}), i.e. neighbors of neighbors.
This rewiring mechanism assumes only local interactions, what is intuitive
for every-day use of language. It was shown that social networks are characterized
by high value of the clustering coefficient
\cite{newman2003social,dorogovtsev2002evolution,foster2011clustering,palla2007quantifying}.
This rewiring mechanism increases
the value of the clustering coefficient by the definition.
Parameters $F$ and $q$ reflects the diversity of language. One trait can stand for a
vocabulary in a given field. Than, different values of $q$ indicate different words
used to describe the same objects.

It was shown that the model defined as above displays three significantly different phases \cite{raducha2017coevolving}.
In the first phase, for small values of $q$, we observe death of most of the dialects.
In this phase, when the system reaches the final configuration almost all agents
speak the same language, and the graph is connected.
In the second phase the network disintegrates into many small components, each with
a different language. Society is polarized and different clusters
use different languages. In the third phase a partial recombination occurs,
but the number of languages increases further, resulting in
existence of links between
individuals speaking different languages. For that reason, the two first phases
are more suitable for the explanation of the language change. Additionally,
it is a reasonable assumption that languages can vary to a finite extent.

Despite the fact that this simple usage-based model of language
manages to capture the essence of social interactions,
its interpretation considering languages was abandoned
after very first publication \cite{axelrod1997dissemination}, due to the contradiction with the empirical data.
Anthropological study of Solomon Islands in the late 70's \cite{terrell1977human} showed that
the number of languages functioning on an island grows with the size of the island.
As noted in the original paper, results of the first model defined on a static square lattice were exactly opposite --
the number of domains was decreasing with increasing size of the lattice. Moreover,
the first adaptive model \cite{sanmiguel2007}, taking into account coevolution of the nodes' states
and the topology of the network, did not solve this issue -- the number of domains
was approximately constant for different sizes of the network.

\begin{figure}[h]
\centering
  \includegraphics[width=0.4\linewidth]{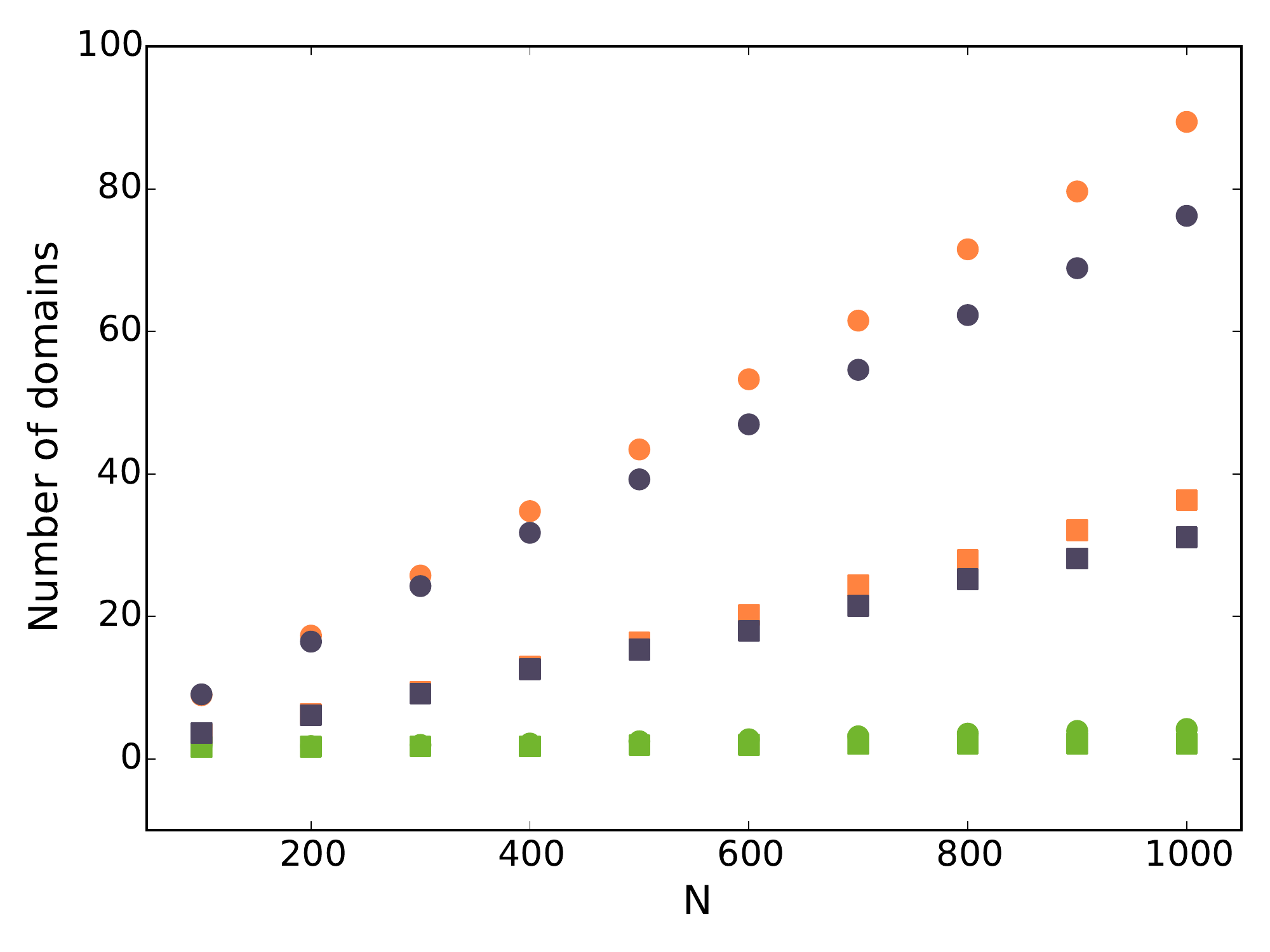}
    \includegraphics[width=0.4\linewidth]{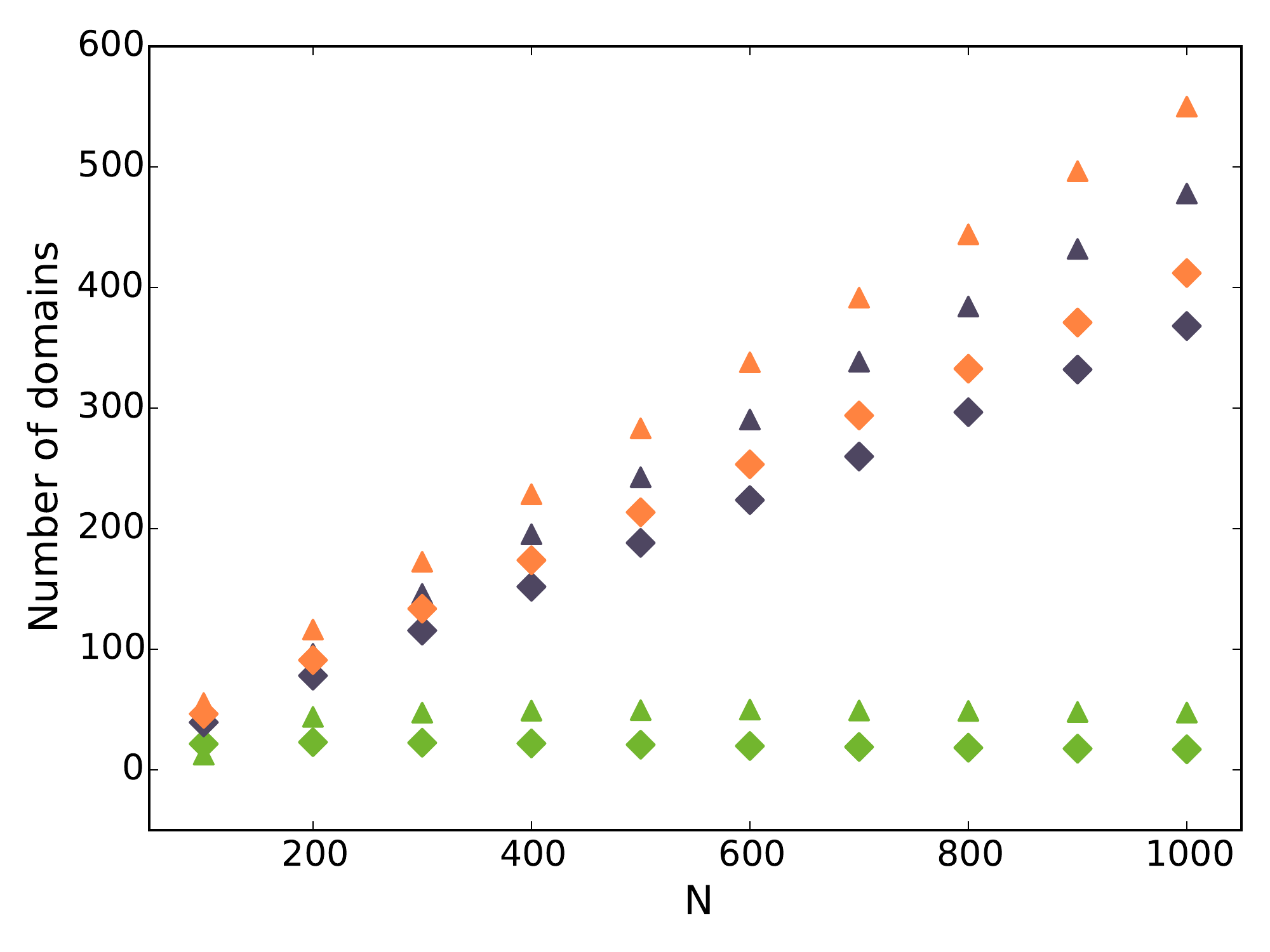}
\caption{Number of domains as a function of the network size $N$
for the uniform model (purple), the model with preferential attachment (orange),
and the model form \cite{sanmiguel2007} (green), for $\langle k \rangle=4$, $F=3$, $q=2$ (squares),
$q=5$ (circles), $q=50$ (diamonds), and $q=100$ (triangles), averaged over 400 realizations.
\label{fig:domains}}
\end{figure}

In FIG. \ref{fig:domains} we analyze behavior of two variants of the model -- local rewiring
with a uniform probability and local rewiring with a preferential attachment. It is clear that
the number of domains, indicating number of languages, increase with the system size.
This result is qualitatively consistent with the empirical data for Solomon Islands given in \cite{terrell1977human}.
It is worth noting that this dependency is also valid, yet weaker, for different models described
in \cite{raducha2017coevolving}, but only for a certain range of values of the parameter $q$.

\begin{figure}[h]
\centering
  \includegraphics[width=0.5\linewidth]{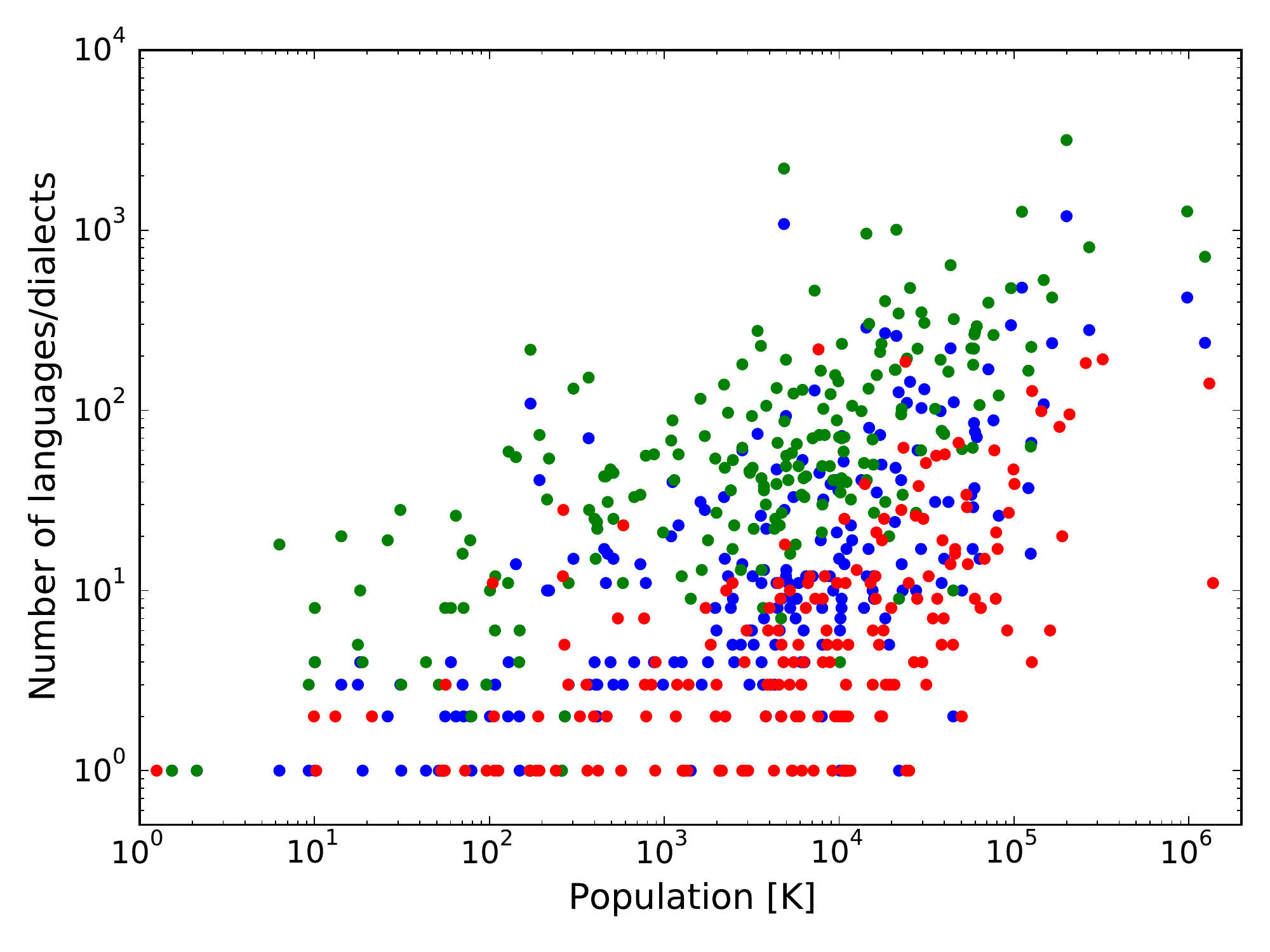}
\caption{Empirical dependence of the number of languages on the size of a population.
Red dots represent languages from \cite{wals2013} and population sizes for 2015 \cite{united2015world}.
Blue dots represent languages from \cite{grimes1996ethnologue} and population sizes for 1996 \cite{united2015world},
green dots represent all dialects from the latter source.
\label{fig:languages}}
\end{figure}

\begin{figure}
\centering
  \includegraphics[width=0.7\linewidth]{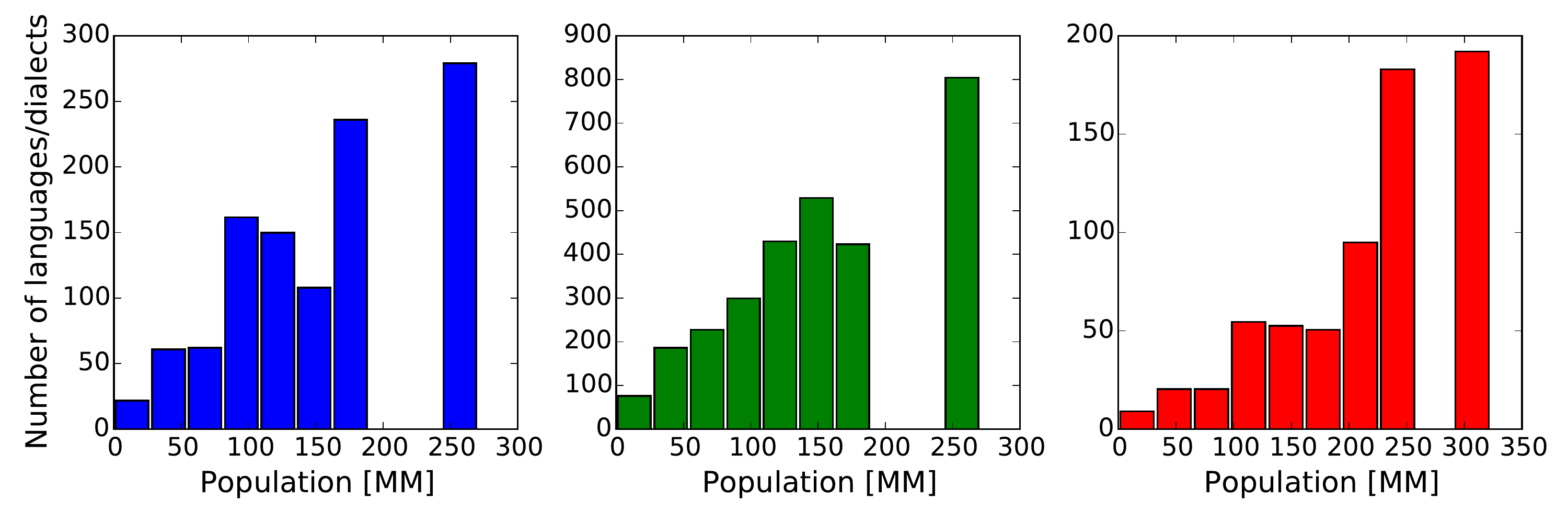}
\caption{Dependence of the number of languages on the size of a population
-- aggregated data from FIG. \ref{fig:languages}, colors preserved.
Height of a bar indicates the average number of languages in countries
with a population lying within the bar.
China, India, Indonesia, and Papua New Guinea are excluded.
\label{fig:languages_hist}}
\end{figure}

Based on our findings, we should expect larger number of languages for countries
with bigger population. To validate this prediction we analyze two databases. The first one form 1996
consisting information about 6866 languages and their 9130 dialects from 209 different countries
\cite{grimes1996ethnologue},
and the second one from 2013 (regularly updated) consisting information about 2679
languages in 188 countries \cite{wals2013}. In FIG. \ref{fig:languages}
we plot the number of languages against the size of a population for countries from six continents.
The trend seems to be increasing in every example, but fluctuations darken the picture.
Obviously, language diversity on a scale of continents is driven not only by social interactions.
There are many factors influencing the linguistic structure of the society, for example
language policy and legislation, colonization, border changes, demolition of the population during wars
or epidemics, compulsory resettlement etc. Nevertheless, we expect our findings to hold
on average. To eliminate fluctuations we aggregate data for consecutive intervals.
Results are shown in FIG.~\ref{fig:languages_hist},
excluding, for the sake of clarity, four countries that have
either the population size (China, India) or the number of languages
(Indonesia, Papua New Guinea) grater by almost order of
magnitude from the others. We obtain growing number of languages
with the population size for both databases. Moreover, this dependency is even more pronounced
in the data set of dialects.
Again, results of the simulations are qualitatively consistent with the empirical data.

In our study we showed that even complex description of nodes' states
in social networks is not sufficient to explain real-world phenomena.
Furthermore, even sophisticated dynamics of states can be not enough
when the structure of the network is divergent to empirical examples.
Topology and its transformations are crucial in a proper description of the language change
due to social interactions. In this field, models with only local rewiring, leading to
high clustering and frequent disintegration, most accurately reproduce empirical data.
Here we have taken steps towards understanding the process of the language change
and its foundations, although its full structure is undiscovered. Nevertheless,
comprehensive model of language should take into account proper dynamics
of the network topology.

\section{Methods}\label{section:methods}

\textbf{Algorithm.}
The model we use is described in details in \cite{raducha2017coevolving}.
We start every simulation with a random graph with $N$ vertices, each representing one agent.
We set the number of links $M$ to obtain a certain value of the average degree $\langle k \rangle$.
Every node $i$ is described by a vector of traits $\sigma_i = (\sigma_{i, 1}, \sigma_{i, 2}, ..., \sigma_{i, F})$.
Every trait can initially adopt one of $q$ discrete values $\sigma_{i, f} \in \{1, 2, ..., q\}, f = 1, 2, ..., F$,
what gives $q^F$ possible different states. At the beginning, we draw a set of $F$ traits
for each node with equal probability for every value form $1$ to $q$.
Then, every time step consists of following rules:
\begin{enumerate}
\item Draw an active node $i$ and one of its neighbors $j$.
\item Compare vectors $\sigma$ of chosen vertices and determine the number $m$
of identical traits (overlap), such that $\sigma_{i, f} = \sigma_{j, f}$.
    \begin{itemize}
    \item If all traits are equal i. e. $m = F$, nothing happens.
    \item If none of the traits are equal $m = 0$, disconnect the edge $(i, j)$
    from node $j$, draw a new node $l$, and attach a link to it, creating an edge $(i, l)$.
    \item In other cases, with probability equal $m/F$ the positive interaction occurs, in which we
    randomly select one of not-shared traits~$f'$ (from among $F - m$) and the active node
    $i$ adopts its value from the node $j$, i. e. $\sigma_{i, f'} \to \sigma'_{i, f'} = \sigma_{j, f'}$.
    \end{itemize}
\item Go to the next time step.
\end{enumerate}
The method of selecting new neighbors is crucial. We allow to create a new connection
only within a set of nodes distant by two edges (neighbors of neighbors).
Multiple connections and auto-connections are prohibited.
We analyze two possibilities: uniform probability for every node in the set, and preferential attachment with
probability $P(i) \sim ( k_i + 1 )^2$. Simulation is ran until frozen configuration is obtained
or thermalization is reached.
In order to describe behavior of the system we use several quantities and coefficients, which are defined as follows.

\textbf{Component $s$}: two vertices $i$ and $j$ belong to the same component $s$, if they are connected,
or vertex $k$ exists such that vertex $i$ belongs to the same component as vertex $k$ and
vertex $k$ belongs to the same component as vertex $j$. Then, by the largest component of the network
we mean the biggest connected subgraph of the network.

\textbf{Domain $d$}: two vertices $i$ and $j$ belong to the same domain $d$, if they are connected
and share all traits $\sigma_{i} = \sigma_{j}$,
or vertex $k$ exists such that vertex $i$ belongs to the same domain as vertex $k$ and
vertex $k$ belongs to the same domain as vertex $j$.
By definition, a given domain cannot exceed the size of the component it shares nodes with.
On the other hand,  the number of components cannot be superior to the number of domains.

\textbf{Local clustering coefficient $c_i$}: for undirected graphs it can be defined as the number
of connections between neighbors of the node $i$ divided by $k_i (k_i -1)/2$, i.e. the number of links that could
possibly exist between them.

\textbf{Global clustering coefficient $C$}: it is defined as three times the number of triangles in the network
divided by the number of connected triplets of vertices (one triangle consists three connected triplets).

\textbf{Average path length $\langle l \rangle$}: it is the shortest distance between two vertices,
averaged over all pairs of vertices in the network. If there is no path between two vertices
(network has many components), this pair is not taken into account.

\textbf{Data Availability.}
The data about languages in different countries that support the findings of this study 
are available in two online databases: Ethnologue \url{www.ethnologue.com/13/names} \cite{grimes1996ethnologue}
and WALS \url{www.wals.info} \cite{wals2013}.
The data about population sizes that support the findings of this study 
are available from United Nations World Population Prospects \url{https://esa.un.org/unpd/wpp} \cite{united2015world}


\begin{acknowledgments}
The authors would like to thank Mateusz Wilinski for discussions and corrections.

%
%
%
%
%
%
%
%
\end{acknowledgments}



\end{document}